\documentclass{PoS}

\title{SFXTs as best candidate counterparts of unidentified transient MeV-GeV sources: \\the test case of IGR J17354$-$3255/AGL J1734$-$3310}

\ShortTitle{the test case of IGR J17354$-$3255/AGL J1734$-$3310}

\author{\speaker{V. Sguera}\thanks{VS is very grateful to Andrea Bulgarelli and the AGILE team for sharing the results on AGL J1734-3310 
before publication. The italian authors acknowledge the ASI financial support via grant ASI-INAF I/033/10/0 and I/009/10/0}\\
        INAF, Istituto di Astrofisica Spaziale e Fisica Cosmica, Via Gobetti 101, I-40129 Bologna, Italy\\
        E-mail: \email{sguera@iasfbo.inaf.it}} 

\author{A. Bazzano\\
       INAF, Istituto di Astrofisica Spaziale e Fisica Cosmica, Rome, Italy\\
}
\author{A. J. Bird\\
       School of Physics and Astronomy, University of Southampton, UK\\
}
\author{S. P. Drave\\
School of Physics and Astronomy, University of Southampton,UK\\
}

\abstract{In the last few years Fermi and AGILE observations have indicated the existence of a possible population of transient MeV-GeV sources located on the Galactic plane and characterized by fast flares lasting only a very few days. Notably, no blazar-like counterparts are known within their error boxes so they could represent a completely new class of Galactic transient high energy emitters. The task of identifying their counterparts at lower energies remains very challenging. Despite this difficulty, INTEGRAL observations have provided intriguing hints that reliable candidate counterparts for these unidentified MeV-GeV transients could be found among the members of the recently discovered class of Supergiant Fast X-ray Transients (SFXTs). In this context, to date the best test case is represented by the association between the two sources IGR J17354$-$3255 and AGL J1734$-$3310. We will discuss their possible physical link and implications stemming from this association.}

\FullConference{The Extreme and Variable High Energy Sky - extremesky2011,\\
		September 19-23, 2011\\
		Chia Laguna (Cagliari), Italy}

\begin{document}

\section{Introduction}
The field of high energy astronomy is relatively young. Breakthrough results have been obtained only 
in the last twenty years thanks to satellites  carrying instruments whose survey capabilities  unveiled
the extreme richness of objects at hard X-rays (E$>$20 keV, e.g. INTEGRAL/IBIS, Swift/BAT) 
as well as gamma-rays (E$>$100 MeV, e.g CGRO/EGRET, AGILE/GRID, Fermi/LAT). 
Interestingly, the great majority of such objects  are still unidentified, with no firmly established
counterparts at lower energies. Their identification  is one of the great challenges of current 
high energy astronomy, it could leave some room for novel and unexpected discoveries. In this context, recent   
AGILE/GRID and Fermi/LAT observations have indicated the existence of a possible population of fast transient MeV-GeV sources 
located on the Galactic plane and characterized by flares lasting only a very few days 
(e.g. Hays et al. 2009, Bulgarelli et al. 2009, Chen et al. 2007). Notably, no blazar-like counterparts are known  within their error boxes so 
they could represent a completely new class of Galactic fast high energy transients. The task of identifying their counterparts at lower 
energies remains very challenging, mainly because of their often  large error circles (e.g radii typically from 10 arcmin  to 0.5 degrees). 
INTEGRAL/IBIS is particularly suited to search for reliable  best candidate counterparts thanks to i) a large field of view 
(FoV) which ensure a total coverage of the gamma-ray error box ii) a good angular resolution which is crucial to disentangle 
the hard X-ray emission of different sources in crowded fields such as those on the Galactic Plane iii) a good sensitivity above 20 keV. 
In particular, recent INTEGRAL results provided intriguing hints that reliable best 
candidate counterparts could be found among the members of the recently  discovered class of Supergiant Fast X-ray Transients 
(Sguera et al. 2009, Sguera et al 2011, Sguera 2009).

Supergiant Fast X-ray Transients (SFXTs) are a new class of High Mass X-ray Binaries (HMXBs) 
mainly unveiled thanks to INTEGRAL observations of the Galactic plane. They host a massive OB supergiant star as 
companion donor (Negueruela et al. 2006) and display X-ray flares lasting from a 
few hours to a few days (Sguera et al. 2005, 2006). It is generally assumed that all
SFXTs host a neutron star as compact object  because their broad band X-ray spectra (0.2--100 keV)
strongly resemble those of accreting X-ray pulsars in HMXBs, i.e. absorbed cutoff power law shape (e.g. Sidoli et al. 2009). 
This idea is supported by the detections of X-ray pulsations in some systems (e.g. Sguera et al. 2007). 
The typical  dynamic range of classical SFXT is 10$^{3}$--10$^{5}$, however some systems 
show a lower value of $\sim$10$^{2}$ and so they have been named as intermediate SFXTs (Sidoli et al. 2011,  Clark et al.  2010, 
Walter \& Zurita 2007). The physical mechanism driving the  peculiar X-ray behaviour of SFXTs  is still unclear and highly debated. 
Several different models  have been proposed in the literature (see Sidoli 2009  for a review). 

We note that in principle SFXTs have all the "ingredients" to possibly be MeV-GeV emitters 
since they host a compact object (i.e. neutron star) as well as a bright and
massive OB star which could act as source of seed photons (for the Inverse Compton emission) and target nuclei (for
hadronic interactions).  In this respect it is important to point out that in the last few years a few classical HMXBs, 
having the same "ingredients" of SFXTs in terms of compact object and companion stellar donor, have been firmly detected at 
MeV-TeV energies as persistent and variable sources (e.g. LS 5039 and LS I +61 303, Paredes 2008, Hill et al. 2010) or as fast flaring transients 
(e.g. Cyg X-3 and Cyg X-1, Tavani et al. 2009, Sabatini et al. 2010), providing evidence that particles can be efficiently accelerated 
to very high energies in HMXBs. At odds with such gamma-ray binaries, the eventual MeV-GeV emission from 
SFXTs must be in the form of fast flares and  should be expected only 
for a very small fraction of time (i.e. from few hours to few days), making very difficult their detection.
Despite this drawback, some observational evidences have been recently  reported in the literature on SFXTs
as best candidate counterparts of unidentified transient MeV-GeV sources located on the Galactic Plane 
(Sguera et al. 2009, Sguera et al 2011, Sguera 2009). These evidences are merely based on intriguing hints such as a spatial correlation and  
a common transient behaviour on  similar, though as yet not simultaneous, short time scales. 
This scenario is also supported from an energetic  standpoint by a theoretical model based in the microquasar 
accretion/jet  framework (Sguera et al. 2009). The so far proposed associations represent an important first step towards 
obtaining reliable candidates on which to concentrate further efforts to obtain quantitative proofs for 
a real physical association.  In this respect, so far, the best test case is represented by the association
between the two sources IGR J17354$-$3255 and AGL J1734$-$3310. 

\section{IGR J17354$-$3255/AGL J1734$-$3310 as test case}
IGR J17354$-$3255 is an unidentified hard X-ray transient discovered by INTEGRAL in 2006 during 
an outburst having average flux of $\sim$ 18 mCrab (20--60 keV) and unconstrained duration (Kuulkers et al. 2006, 2007). 
There are some good reasons to believe that IGR J17354$-$3255 is a HMXB hosting a neutron star as compact object 
(D'Ai et al. 2011, Tomsick et al. 2009): i) the soft X-ray spectrum (0.2--10 keV) is well represented by a 
rather hard power law ($\Gamma$ $\sim$ 0.5) modified by intrinsic absorption (N$_H$ $\sim$  7$\times$10$^{22}$ cm$^{-2}$), 
ii) a cutoff energy around 20 keV  is requierd in the power law broad band X-ray spectrum 0.2--100 keV,
iii) the $\sim$ 8.4 days periodicity in the long term hard X-ray light curve  which is very likely the orbital period 
of the binary system, iv) a bright 2MASS infrared counterpart.

With the aim of investigating the temporal hard X-ray behaviour of IGR J17354$-$3255, which is largely unknown,
we performed a detailed study of the 18--60 keV INTEGRAL/IBIS  long-term light curve on Science Window timescale (ScW, $\sim$ 2000 s)
for a total on source time of $\sim$ 10 Ms. We characterise IGR J17354$-$3255 as a weak persistent hard X-ray source spending 
a major fraction of the time during an out-of-outburst state with an average 18--60 keV flux of 
$\sim$ 1.1 mCrab. This is occasionally interspersed with fast hard X-ray flares (duration from
a few hours to a few days) for a dynamic range in the interval 20--200. A total of 16 hard X-ray flares have been detected by INTEGRAL/IBIS 
over a total exposure of $\sim$ 115 days though not in sequence, as listed in Table 1.
The high dynamic range  of IGR J17354$-$3255  is also confirmed in the softer X-ray band (0.2--10 keV) 
thanks to archival Swift/XRT observations from which we inferred a lower limit value of $\sim$ 310
from non detection (3$\sigma$ upper limit of 2.8$\times$10$^{-13}$ erg cm$^{-2}$ s$^{-1}$) to flaring activity (8.7$\times$10$^{-11}$ 
erg cm$^{-2}$ s$^{-1}$).  When active and detected by Swift/XRT, the source was also strongly variable 
by a factor of $\sim$ 50 on time-scales of only a few hours. 

\begin{table}
\begin{center}
\caption {Summary of all INTEGRAL/IBIS detections of hard X-ray flares from IGR J17354$-$3255. The table lists the date 
of their peak emission, approximate duration  and significance detection of the entire flaring activity,  X-ray flux  at the peak, power law photon index with   
$\chi^{2}_{\nu}$ and  d.o.f., and finally reference to the discovery paper of each flare, i.e. (1) this work and  (2) Kuulkers et al 2006; $\ddagger$ = upper limit on the duration,  $\star$  = lower limit on the duration.}
\label{tab:main_outbursts} 
\small
\begin{tabular}{ccccccc}
\hline
N. & peak-date    &  duration    & sig    &  peak-flux       & $\Gamma$                   & ref \\
    &  (MJD)     &   (hours)     &  $\sigma$        & (18--60 keV)     &($\chi^{2}_{\nu}$, d.o.f.)  &      \\
   &            &                &                  &   (mCrab)        &                            &      \\
\hline                                                                            
1  & 52741.5  & $\sim$65$\ddagger$  & 9.0     & 25$\pm$5        & 2.0$^{+0.7}_{-0.7}$ (1.2,15)                   & 1   \\
2  & 53051.9 &$\sim$0.5             & 5.5   & 108$\pm$20        & 2.2$^{+1.0}_{-2.0}$ (1.2,15)        & 1   \\
3  & 53114.9  &$\sim$10             & 5.5   & 35$\pm$12       & 2.3$^{+2.0}_{-2.0}$ (1.2,15)             & 1   \\
4  & 53452.4  &$\sim$0.5            & 4.2   & 25$\pm$6        &                           & 1   \\
5  & 53602.9  &$\sim$0.5            & 4.4   & 35$\pm$8        &                           & 1   \\
6 & 53794.6  &$\sim$0.5             & 5.0     & 25$\pm$5        & 2.6$^{+2.0}_{-2.5}$ (1.3,15)    & 1   \\
7 & 53813.8  &$\sim$60$\ddagger$    & 5.6   & 27$\pm$6        & 2.2$^{+1.6}_{-1.6}$  (0.9,15)       & 1   \\
8 & 53829.8  &$\sim$36              & 6.6   & 30$\pm$4        & 2.6$^{+0.8}_{-0.8}$  (0.8,15)      & 1   \\
9 & 53846.2  &$\sim$5$\star$        & 8.0   & 28$\pm$6        & 2.0$^{+0.7}_{-0.7}$   (0.8,15)      & 2   \\
10 & 53975.3  &$\sim$3              & 4.6   & 32$\pm$7        &                           & 1   \\
11 & 53999.7  &$\sim$1.5            & 5.2   & 30$\pm$7        & 2.1$^{+1.2}_{-1.2}$ (1.1,15)      & 1   \\
12 & 54012.6  &$\sim$5              & 4.2   & 40$\pm$9        &                          & 1   \\
13 & 54340.2  &$\sim$5              & 6.5   & 25$\pm$7        & 2.5$^{+2.0}_{-2.0}$  (1.01,15)      & 1   \\
14 & 54345.5  &$\sim$63             & 7.0  & 21$\pm$6        & 3.2$^{+1.0}_{-1.0}$ (1.2,15)       & 1   \\
15 & 54539.1  &$\sim$25             & 6.2   & 21$\pm$5        & 2.1$^{+1.5}_{-2.0}$ (1.2,15)          & 1   \\
16 & 54547.8  &$\sim$3$\star$       & 5.7   & 30$\pm$6        & 2.9$^{+1.0}_{-1.0}$ (0.9,15)        & 1   \\
\hline
\hline  
\end{tabular}
\end{center}
\end{table} 

Our detailed INTEGRAL/IBIS timing analysis strongly confirmed  the 8.4 days  orbital period. 
The occurrence of all the outbursts detected by INTEGRAL/IBIS  is consistent with the  orbital phase  of periastron
passage of the likely neutron star compact object during its  orbit around the companion donor star. 
In addition, we also note that  the shape of the orbital profile is rather smooth and this cannot be explained by 
the sixteen individual outbursts detected by INTEGRAL/IBIS. Assuming a source distance of 8.5 kpc, these outbursts all have X-ray luminosities 
in excess of 10$^{36}$ erg s$^{-1}$ and thus should represent the most luminous outburst events. Hence we would not expect these events to 
define the orbital emission profile over the extent of the long-baseline observations. Instead we attribute the
shape to the lower level X-ray emission that is below the instrumental sensitivity of INTEGRAL/IBIS in an individual ScW (i.e. $\sim$ 10 mCrab).
However when the whole data set, covering about 300 orbital cycles of 8.4 days, is folded this emission sums to a significant detection 
and reveals the smooth profile.

All  our reported  findings on IGR J17354$-$3255 are highly indicative of a HMXB hosting 
a supergiant star as companion donor (SGXB), however the inferred dynamic ranges both at hard (20--220) and soft (310) X-rays 
are significantly  greater than those of classical persistent SGXB systems ($<$20). We suggest that IGR J17354$-$3255 is 
an intermediate SFXT, much like other similar cases reported in the literature (e.g. Walter \& Zurita 2007; Clark et al. 2010, Sidoli et al. 2011).

Interestingly, we note that IGR J17354$-$3255 is the only hard X-ray source located inside 
the error circle of the unidentified transient MeV-GeV source AGL J1734$-$3310 (E$>$100 MeV).
AGL J1743$-$3310 was discovered by the AGILE gamma--ray satellite on 2009 April 14 during a flare  lasting 
only 1 day and detected with a significance of  4.8$\sigma$ (Bulgarelli et al. 2009). 
After the discovery of AGL J1734$-$3310, extensive searches for further flaring 
gamma-ray emission have been carried  out by the AGILE team (Bulgarelli et al. in preparation, private communication). 
As a result, 9 additional MeV-GeV flares have been discovered in the AGILE data archive 2007--2009. They have a similar duration (about 1 day)
and significance detection in the range (3--5)$\sigma$. This clearly shows that AGL J1734$-$3310 is a recurrent transient MeV-GeV source. 
The significance of the sum of all the flares detected by AGILE is 7.3$\sigma$ with a 95\%  statistical 
and systematic positional error radius of 0.46 degrees. 
Fig. 1 shows the 18--60 keV INTEGRAL/IBIS  significance mosaic map ($\sim$ 8 Ms exposure) of the sky region 
surrounding IGR J17354$-$3255 with superimposed the positional uncertainty of AGL J1734$-$3310 (green circle). 
Clearly, IGR J17354$-$3255 is the only hard X-ray source  detected inside the AGILE error circle, 
the same holds in the softer X-ray band (3--10 keV) from a JEM--X deep mosaic ($\sim$ 650 ks exposure). 
For the sake of completeness, we note that in the surroundings of the AGILE error circle there are  a few more MeV-GeV
sources (see figure 1): i) 3EG J1734$-$3232 is a still unidentified EGRET source, it is likely variable as 
suggested by the value of its  variability index I (Han \& Zhang 2005). This, together  with the spatial match, 
likely suggest that 3EG J1734$-$3232  and AGL J1734$-$3310 could be the same source, i.e. IGR J17354$-$3255
ii) 2FGL J1732.5$-$3131 is a firmly identified gamma-ray pulsar and this unambiguously excludes its association with 
3EG J1734$-$3232 iii) 2FGL J1737.2$-$3213 is an unidentified gamma-ray source, it is not variable and 
it is likely associated with a supernova remnant or pulsar wind nebula as listed in the  latest Fermi catalog (Abdo et al. 2011),  because of this 
we likely exclude its association with AGL J1734$-$3310/3EG J1734$-$3232 
iv) 2FGL J1731.6$-$3234c is still unidentified, however as reported in the latest Fermi catalog it is 
found in a region with possibly incorrected diffuse emission. As such, its position and even existence may not be reliable, i.e. 
it could be a fake source potentially confused with interstellar emission.
\begin{figure}
\begin{center}
\includegraphics[width=.8\textwidth]{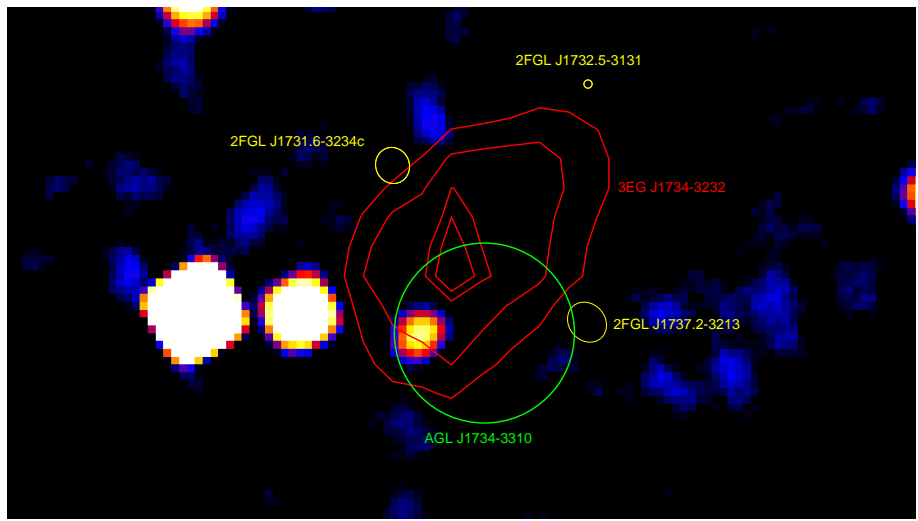}
\caption{The INTEGRAL/IBIS mosaic significance map (18--60 keV, $\sim$ 8 Ms exposure time) of the sky region including IGR J17354$-$3255. 
The other two bright sources detected in the field are the LMXBs GX 354$-$0 and 4U 1730$-$335.
The green error circle represents the MeV-GeV source AGL J1734$-$3310 and the red contours (from 50\%  to 99\%) refer to 
3EG J1734$-$3232. The gamma-ray sources detected by Fermi/LAT are indicated by means of yellow circles} 
\end{center}
\end{figure}

It is interesting to note that the spatial correlation  between AGL J1734$-$3310 and IGR J17354--3255 is also
supported by a similar flaring nature  on similar short timescales.  This is because we propose the intermediate SFXT
IGR J17354$-$3255 as best candidate counterpart of AGL J1734$-$3310, to date. 
Although such proposed association is merely based on intriguing hints, it represents an important first step towards 
obtaining reliable test cases on which to concentrate further efforts to obtain 
quantitative proofs for a real physical association. In this respect, further AGILE and INTEGRAL studies of IGR J17354$-$3255/AGL J1734$-$3310 
are under way with the aim of searching for i) the 8.4 days periodicity in the AGILE gamma-ray data ii) 
flares from IGR J17354$-$3255 simultaneously detected both by  AGILE and INTEGRAL. If fully confirmed, the implications of SFXTs producing 
MeV-GeV emission are huge, both theoretically and observationally, and would add a further extreme characteristic 
to this already extreme class of fast transient sources.

\end{document}